\documentclass[final,3p,times]{elsarticle}

\usepackage[utf8]{inputenc}
\usepackage[T1]{fontenc}

\usepackage{amssymb}
\usepackage{amsmath}

\tnotetext[cernpreprint]{CERN-TH-2026-215}

\journal{Journal of Subatomic Particles and Cosmology}

\begin{document}

\begin{frontmatter}



\title{Spectator shadowing as the source of the breaking of NCQ scaling at RHIC-FXT and FAIR energies}

\author[Cern,Duke]{Tom Reichert}
\author[CTU]{Iurii Kaprenko}

\affiliation[Cern]{
    organization={Theoretical Physics Department, CERN},
    postcode={1211},
    city={Geneva 23},
    country={Switzerland}
}

\affiliation[Duke]{
    organization={Department of Physics, Duke University},
    city={Durham},
    state={NC},
    postcode={27708},
    country={USA}
}

\affiliation[CTU]{
    organization={Faculty of Nuclear Sciences and Physical Engineering, Czech Technical University in Prague},
    addressline={B\v{r}ehov\'a 7},
    postcode={11519},
    city={Prague 1},
    country={Czech Republic}
}

\begin{abstract}
The apparent scaling of the elliptic flow of identified hadrons with the number of constituent quarks indicates the presence of quark coalescence, and in turn suggests the presence of a deconfined state in heavy-ion collisions. Recent measurements at RHIC found a breaking of NCQ scaling at low energies and suggested that this marks the onset of partonic collectivity at higher energies. Here, we present a new framework embedding the effect of spectator shadowing into the flow scaling relations. Using a toy model with an idealized quark coalescence source and a ballistic Glauber model for the bypassing spectator, we demonstrate that the observed breaking is an effect of spectator shadowing. 
\end{abstract}



\begin{keyword}
Heavy-ion collision\sep Elliptic flow\sep Constituent quark number scaling\sep Quark coalescence\sep Spectator shadowing


\end{keyword}

\end{frontmatter}

\section{Introduction}
Strongly interacting matter, described by Quantum Chromodynamics (QCD), is studied in heavy-ion collisions over a wide range of beam energies. At high temperatures and vanishing baryon chemical potential, lattice QCD predicts a crossover transition to the Quark-Gluon Plasma (QGP) \cite{Cabibbo:1975ig,Borsanyi:2010bp,Bazavov:2011nk,Bazavov:2017dus}. At finite baryon chemical potential, the QCD phase structure remains less constrained and is currently explored by low-energy programs such as STAR-FXT and the future CBM experiment \cite{Sorensen:2023zkk,Agarwal:2025ezo,Messchendorp:2025men}.

One possible signal for partonic collectivity, and thus the presence of a partonic stage, is constituent quark number scaling of the elliptic flow, $v_2^h(p_\mathrm{T}^h)=N_q v_2^q(p_\mathrm{T}^q)$ with $p_\mathrm{T}^h=N_q p_\mathrm{T}^q$. While such scaling has been observed at several collision energies \cite{STAR:2012och,STAR:2013cow,STAR:2013ayu,ALICE:2014wao,STAR:2015rxv,HADES:2022osk}, its interpretation remains puzzling \cite{Lu:2006qn} and at few-GeV collision energies it is further complicated by spectator shadowing \cite{Wang:2024ktk,Reichert:2024ayg}. The spectators remain close to the fireball during the course of the collision and absorb transverse momentum, thereby modifying the measured azimuthal particle distribution and generating negative elliptic flow. This effect is well known and commonly used to infer the pressure present during the collision, i.e. the equation of state of compressed nuclear matter, \cite{Sorensen:2023zkk,Danielewicz:2002pu,Oliinychenko:2022uvy,Steinheimer:2025hsr}.

Recent STAR-FXT measurements report a breaking of NCQ scaling below $\sqrt{s_\mathrm{NN}}\approx 4.5$ GeV \cite{STAR:2021yiu,STAR:2025owm}. In this work, we expand the model we have developed in Ref. \cite{Reichert:2026dsb} and propose a method to subtract the spectator contribution from the measured elliptic flow. The method is based on a Fourier decomposition of the transverse shadowing strength, expressed through an absorption probability or optical depth, and allows one to reconstruct the azimuthal structure of the emission source. We then show that concluding a ``disappearence of partonic collectivity'' from measured data is not straightforward.

\section{NCQ scaling with spectator shadowing}
In a quark coalescence picture, hadrons are formed from nearby quarks in phase space. In the most general case, the hadron's Wigner function has a finite width and the space-momentum correlations of the source will affect the coalescence calculation \cite{Molnar:2004rr,Pratt:2004zq}. We adopt the typical assumption that space-momentum correlations are absent and that the Wigner function is narrow. This leads to the usual proportionality
\begin{align}
    \mathrm{d}N^h(p_\mathrm{T}) &\propto \left[ \mathrm{d}N^q(p_\mathrm{T}/N_q) \right]^{N_q},
\end{align}
with $N_q=2$ for mesons and $N_q=3$ for baryons. The azimuthal distribution of the quarks can be expanded as a Fourier series
\begin{align}
    f^q(\phi) \propto 1+2\sum_{n=1}^{\infty} v_n^q \cos(n\phi).
\end{align}
Consequently, in the coalescence limit, the hadronic distributions are given by powers of the quark distribution $F_M(\phi) \propto \left[f^q(\phi)\right]^2$ and $F_B(\phi) \propto \left[f^q(\phi)\right]^3$.
Expanding these expressions and collecting terms with equal harmonic gives, to leading order, the well known NCQ scaling relations
\begin{align}
    v_2^M(p_\mathrm{T}) &= 2 v_2^q(p_\mathrm{T}/2), \\
    v_2^B(p_\mathrm{T}) &= 3 v_2^q(p_\mathrm{T}/3),
\end{align}
while higher order corrections mix different partonic harmonics \cite{Nonaka:2003hx,Molnar:2003ff,Kolb:2004gi}. The derivation assumes equal azimuthal distributions for all quark flavors. Extensions to flavor-dependent, transported or produced quarks, or treatment of imperfect correlations are straightforward and can be found elsewhere \cite{Molnar:2004rr,Pratt:2004zq,Dunlop:2011cf}.

At few-GeV collision energies this picture is modified by spectator shadowing. In contrast to high-energy collisions, the spectators do not immediately decouple from the interaction region, but remain close to the expanding fireball during their finite passing time, $t_\mathrm{pass}\sim(\sqrt{s_\mathrm{NN}})^{-1}$. The measured hadron distribution is therefore not only determined by the source anisotropy, but also by the probability that a hadron escapes without further absorption or rescattering. For a hadron produced at $(t,\mathbf{x})$ and propagating along a classical trajectory, this escape probability can be written as 
\begin{align}
    P_\mathrm{esc}(\phi) &\propto \exp\left[ -\int_t^\infty \mathrm{d}t^\prime \sigma(\sqrt{s}) |v_\mathrm{rel}| \rho(t^\prime,\mathbf{x}^\prime) \right],
\end{align}
which is the on-shell limit of the more general description of Ref. \cite{Knoll:2008sc}. Since the spectator geometry is anisotropic in space, $P_\mathrm{esc}$ is itself angle dependent and can be expanded as
\begin{align}
    P_\mathrm{esc}(\phi) &\propto 1 + 2\sum_{n=1}^{\infty} p_n \cos(n\phi),
\end{align}
where the coefficients $p_n$ encode the harmonic structure of the shadowing strength and the space-momentum correlations of the source and the spectator. Note that the coefficients are species dependent through the interaction cross section and depend on the space-time geometry of the collision.

The experimentally measured azimuthal distributions are therefore given by the product of the source distribution and the escape probability $\mathcal{F}_M(\phi) \propto \left[f^q(\phi)\right]^2 P_\mathrm{esc}(\phi)$ and $\mathcal{F}_B(\phi) \propto \left[f^q(\phi)\right]^3 P_\mathrm{esc}(\phi)$ if one assumes perfect quark coalescence.
The measured harmonic coefficients are obtained from
\begin{align}
    \mathcal{V}_n^h &= \frac{ \int \mathrm{d}\phi \cos(n\phi) \left[f^q(\phi)\right]^{N_q} P_\mathrm{esc}(\phi)}{\int \mathrm{d}\phi \left[f^q(\phi)\right]^{N_q} P_\mathrm{esc}(\phi)} .
\end{align}
To leading order in the anisotropies this gives the simple relation
\begin{align} \label{eq:sNCQ}
    \mathcal{V}_n^h(p_\mathrm{T}) &\simeq N_q v_n^q(p_\mathrm{T}/N_q) + p_n^h(p_\mathrm{T}) ,
\end{align}
showing that the measured hadron flow contains both the intrinsic source anisotropy and the shadowing contribution. In the limit $p_n\rightarrow 0$, which is restored at large collision energies where the spectators decouple instantly, the standard coalescence result of Ref. \cite{Kolb:2004gi} is recovered. It is also clear from Eq. \eqref{eq:sNCQ} that the scaling can be recovered by subtracting the shadowing effect as $\mathcal{V}_n^h(p_\mathrm{T}) - p_n^h(p_\mathrm{T}) \simeq N_q v_n^q(p_\mathrm{T}/N_q)$. 

The full higher order expressions, which contain the mixing between different $v_n^q$ and $p_n$, are given in our previous letter \cite{Reichert:2026dsb}. There we have derived the expanded expressions for a shadowed quark coalescence source for mesons and baryons. For completeness, we add here the expressions of a shadowed pure hadronic source, i.e. without quark coalescence, up to fourth order in flow and shadowing coefficients. These coefficients arise from expanding $\mathcal{F}_h(\phi) \propto f^h(\phi) P_\mathrm{esc}(\phi)$ for hadron $h$.
\begin{align}
    \mathcal{V}_1^h &= \frac{1}{\mathcal{N}^h} \left[ v_1 + p_1^h \left( 1 + v_2 \right) + p_2^h \left( v_1 + v_3 \right) + p_3^h \left( v_2 + v_4 \right) + p_4^h v_3 \right] \\
    \mathcal{V}_2^h &= \frac{1}{\mathcal{N}^h} \left[ v_2 + p_1^h \left( v_1 + v_3 \right) + p_2^h \left( 1 + v_4 \right) + p_3^h v_1 + p_4^h v_2 \right] \\
    \mathcal{V}_3^h &= \frac{1}{\mathcal{N}^h} \left[ v_3 + p_1^h \left( v_2 + v_4 \right) + p_2^h v_1 + p_3^h + p_4^h v_1 \right] \\
    \mathcal{V}_4^h &= \frac{1}{\mathcal{N}^h} \left[ v_4 + p_1^h v_3 + p_2^h v_2 + p_3^h v_1 + p_4^h \right] \\
    \mathcal{N}^h &= 1 + 2 p_1^h v_1 + 2 p_2^h v_2 + 2 p_3^h v_3 + 2 p_4^h v_4
\end{align}

\section{Toy model}
It is valuable to obtain (semi-)analytic expressions for the shadowing coefficients $p_n^h$ that include the collision energy $\sqrt{s_\mathrm{NN}}$, the spectator passing time $t_\mathrm{pass}$, and the relevant hadronic cross sections. Such coefficients could then be used to correct measured flow data for spectator shadowing and to test whether constituent quark number scaling is present in the underlying emitting source. Ultimately this task has to be addressed by means of a dynamical microscopic model including realistic space-momentum correlations, resonance excitation and feed-down, etc. Here, we employ a toy model to demonstrate the qualitative effect of shadowing and to show that its magnitude is dominant at low collision energies.

We calculate the measurable elliptic flow $\mathcal{V}_2$ at midrapidity, where odd harmonics vanish by symmetry, from a coalescing source with a parametrized partonic elliptic flow and a transverse absorption probability. The parton elliptic flow is chosen as
\begin{align}
    v_2^q(p_\mathrm{T}^q) &= v_{2,\mathrm{max}}^q \tanh\left(p_\mathrm{T}^q/\Lambda\right),
\end{align}
with $v_{2,\mathrm{max}}^q=0.05$ and $\Lambda=0.5$ GeV, following Ref. \cite{Dunlop:2011cf}. This determines the elliptic flow of the mesons and baryons before shadowing simply as $v_2^M = 2 v_2^q$ and $v_2^B = 3 v_2^q$. The shadowing contribution is estimated using a ballistic Glauber model similar to Ref. \cite{Vovchenko:2014gda}. The spectator density is modeled by a Woods-Saxon profile propagating without stopping as
\begin{align}
    \rho(t,r) &= \frac{\gamma\rho_0}{1+\exp\left[(r(t)-R_0)/\sigma_r\right]}, \qquad r(t) = \sqrt{\left(x\mp\frac{b}{2}\right)^2 + y^2 + \gamma^2 \left(z\pm\frac{R_0}{\gamma}\mp\beta t\right)^2} .
\end{align}
All hadrons are assumed to be emitted from a point-like source at the origin at the time of full nuclear overlap, $t_\mathrm{overlap}=R/(\beta\gamma)$. Their transverse velocity at midrapidity is given by $p_\mathrm{T}/m_\mathrm{T}$. The resulting escape probability $P_\mathrm{esc}(\phi)$ determines the shadowing coefficients $p_n^h$ for each hadron species.

\begin{figure}[t]
    \centering
    \includegraphics[width=\columnwidth]{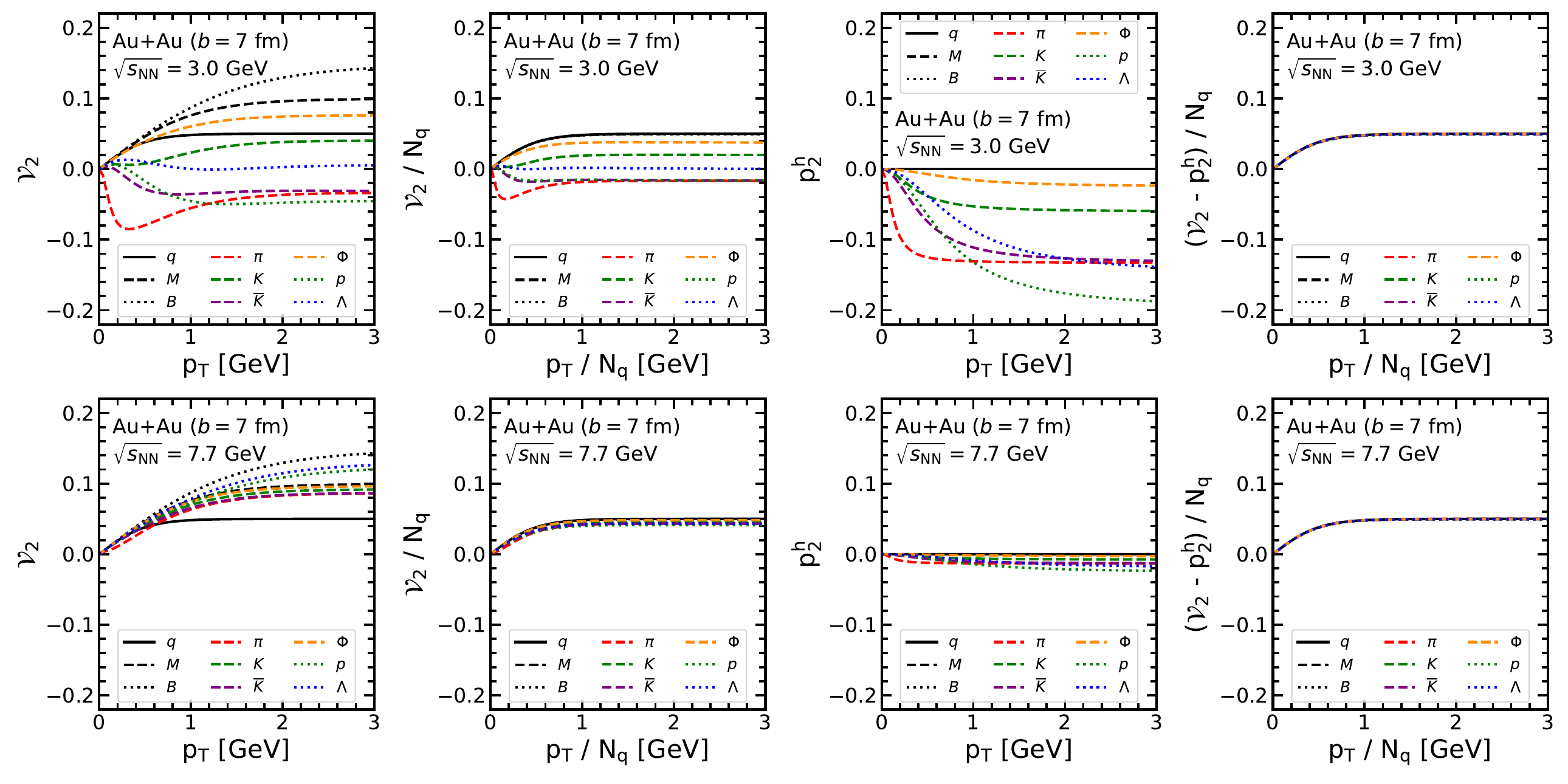}
    \caption{Toy model calculation of the measurable elliptic flow (1st column), the naively scaled measured elliptic flow (2nd column), the shadowing coefficient (3rd column) and the properly scaled elliptic flow (4th column) of $\pi$, $K$, $\overline{K}$, $\phi$, $p$, $\Lambda$ from an idealized quark coalescence source shadowed by a ballistic Glauber model.}\label{fig:toy_model}
\end{figure}

As an illustrative example, we consider peripheral Au+Au collisions at $b=7$ fm at $\sqrt{s_\mathrm{NN}}=3.0$ GeV and $7.7$ GeV. We compare the unshadowed coalescence result to the measurable elliptic flow of $\pi$, $K$, $\overline{K}$, $\phi$, $p$, and $\Lambda$ using constant effective absorption cross sections: $\sigma^\mathrm{eff}_{\pi N}=50 \mathrm{mb}$, $\sigma^\mathrm{eff}_{K N}=12 \mathrm{mb}$, $\sigma^\mathrm{eff}_{\overline{K}N}=50 \mathrm{mb}$, $\sigma^\mathrm{eff}_{\phi N}=5 \mathrm{mb}$, $\sigma^\mathrm{eff}_{pN}=40 \mathrm{mb}$, $\sigma^\mathrm{eff}_{\Lambda N}=30 \mathrm{mb}$.
The pion and antikaon values are chosen much smaller than their free cross sections, since repeated resonance excitation and decay can strongly repopulate their yields at low energies. Figure \ref{fig:toy_model} shows that shadowing is much stronger at $\sqrt{s_\mathrm{NN}}=3.0$ GeV than at $7.7$ GeV, reflecting the longer spectator passing time, as seen in the measurable flow signal in the left column. Consequently, the naive NCQ scaled elliptic flow $\mathcal{V}_2/N_q$ (in the second column) is clearly broken at the lower energy, while it remains approximately preserved at the higher energy. This is qualitatively consistent with the STAR-FXT observation of NCQ scaling breaking at low collision energies \cite{STAR:2021yiu,STAR:2025owm}. The extracted shadowing coefficient $p_2^h$ exhibits a clear hierarchy with the effective hadronic cross section and also depends on the hadron mass through the propagation velocity $v_\mathrm{T} = p_\mathrm{T}/m_\mathrm{T}$, cf. the third column. Subtracting the shadowing contribution before scaling, $(\mathcal{V}_2-p_2^h)/N_q$, reconstructs the unshadowed source elliptic flow and restores the imposed NCQ scaling by construction, as seen in the last column.

We emphasize that this toy model is not intended as a quantitative correction of experimental data. It neglects baryon stopping, transverse expansion, realistic energy and angle dependent cross sections, resonance excitations, the full space-time structure of hadron production and realistic space-momentum correlations. These effects are naturally included in hadronic transport models, which are therefore the appropriate tool to calculate the coefficients $p_n^h$ for realistic comparisons. Such a study will be presented in future work \cite{Reichert:tbp}. However, the toy model demonstrates that spectator shadowing generates a split in elliptic flow with the same qualitative behavior and quantitative magnitude as the breaking of NCQ scaling would.

\section*{Acknowledgements}
The author thanks the SQM2026 organizers. 
We thank Agnieszka Sorensen, Paul Sorensen, Richard Seto, Volodymyr Vovchenko and Volker Koch for fruitful discussions about NCQ scaling during the INT workshop ``The QCD Critical Point: Are We There Yet?'' under event code INT-25-3a that have inspired this study. 
We thank the Institute for Nuclear Theory at the University of Washington for its kind hospitality and stimulating research environment. 
T.R. thanks Steffen Bass for the kind hospitality at Duke University. 
T.R. gratefully acknowledges financial support by the Fulbright U.S. Scholar Program, which is sponsored by the U.S. Department of State and the German-American Fulbright Commission. This article’s contents are solely the responsibility of the author and do not necessarily represent the official views of the Fulbright Program, the Government of the United States, or the German-American Fulbright Commission. 
T.R. gratefully acknowledges support from The Branco Weiss Fellowship - Society in Science, administered by the ETH Z\"urich. IK acknowledges support by the Czech Science Foundation under project No. 25-16877S.

\bibliographystyle{elsarticle-num}
\bibliography{sqm2026_template}

\end{document}